\documentclass[a4paper,12pt]{article}
\usepackage[left=2cm,right=2cm]{geometry}
\usepackage[utf8x]{inputenc}
\usepackage{amsfonts,amsmath,amssymb,bm,eucal}
\usepackage{graphicx}
\usepackage{epstopdf}
\usepackage{subfigure}

\newcommand{\be}{\begin{equation}}
\newcommand{\ee}{\end{equation}}
\newcommand{\bc}{\begin{center}}
\newcommand{\ec}{\end{center}}
\newcommand{\bea}{\begin{eqnarray}}
\newcommand{\eea}{\end{eqnarray}}
\newcommand{\nn}{\nonumber}

\title{\vspace*{-2cm}
\begin{flushright}
\normalsize{BARI-TH/686-2014}
\end{flushright}
\vspace*{1.5cm}Exotic $J^{PC}=1^{-+}$ mesons \\ in a holographic   model of QCD}
\author{L.~Bellantuono$^{a,b}$, P.~Colangelo$^a$ and F.~Giannuzzi$^{a,b}$\\~\\ 
\normalsize\emph{$^a$INFN - Sezione di Bari, via Orabona 4, I-70126, Bari, Italy}\\
\normalsize\emph{$^b$Dipartimento di Fisica, Universit\`a degli Studi di Bari, via Orabona 4, I-70126, Bari, Italy}
}
\date{}

\begin{document}

\maketitle

\begin{abstract}
Mesons with quantum numbers $J^{PC}=1^{-+}$ cannot be represented as simple quark-antiquark pairs. We explore hybrid configurations in the light meson sector comprising a quark, an antiquark and an excited gluon, studying the properties of such states in a phenomenological model inspired by the gauge/gravity correspondence. The computed mass,  compared to the experimental mass of the $1^{-+}$ candidates $\pi_1(1400)$, $\pi_1(1600)$ and  $\pi_1(2015)$, favous $\pi_1(1400)$ as the lightest hybrid state.
An interesting result  concerns the stability of hybrid mesons at finite temperature: they disappear  from the spectral function  (i.e. they melt) at a lower temperature with respect to other states, light vector and scalar mesons, and scalar glueballs.

\vspace*{0.3cm}
\noindent{pacs:
11.25.Tq, 
11.10.Kk, 
11.15.Tk, 
14.40.Rt 
}
\end{abstract}

\vspace*{1cm}
\section{Introduction}\label{sec:intro}
There are combinations of spin, parity and charge conjugation of meson quantum numbers  that cannot be obtained by a quark model description of a quark-antiquark pair with orbital angular momentum $L$ and quark spin $S_q$ and $S_{\bar q}$. This is the case of the $J^{PC}$ ``exotic" combinations $0^{--}$, $0^{+-}$, $1^{-+}$,  etc. Nevertheless,  hadrons with those quantum numbers  were envisaged since the early  high energy theoretical and experimental investigations  \cite{early}, and their search   has continued till   the present days. An argument in favour of the existence of the exotic  resonances is that they could be obtained by  adding a gluonic excitation to the $q \bar q$ pair,  and therefore they represent  a ``hybrid"  $q G\bar q$  state
 with  the unconventional $J^{PC}$ allowed.
Such exotic hybrid configurations should be observed as additional states  in the meson spectrum.

From the experimental side, 
investigations  in the heavy quark sector have reported  of several mesons with nonconventional features; however,  there is no observation (so far)  of states  with exotic quantum numbers \cite{Swanson:2006st}. On the other hand, 
in the light quark sector  there are at least three quite  well established  hybrid candidates: $\pi_1(1400)$, $\pi_1(1600)$ and $\pi_1(2015)$.

The $J^{PC}=1^{-+}$ state $\pi_1(1400)$ has been  reported by the  E852 Brookhaven experiment  \cite{Thompson:1997bs,Chung:1999we}
 in the process $\pi^- p\to \eta \pi^- p$; it was introduced to explain an asymmetry in the $\eta \pi$ angular distribution which could indicate an interference between $\ell=1$ and $\ell=2$ partial waves.  Other analyses are described in Refs. \cite{Abele:1998gn,Abele:1999tf,Salvini:2004gz,Adams:2006sa}.  The measured  mass is $M(\pi_1(1400))= 1354\pm 25$ MeV \cite{Beringer:1900zz}.

$\pi_1(1600)$, again with $J^{PC}=1^{-+}$,  has been  also reported in  $\pi^- p$ interactions \cite{Adams:1998ff,Ivanov:2001rv,Kuhn:2004en,Lu:2004yn,Alekseev:2009aa} with mass  $M(\pi_1(1600))=1662^{+8}_{-9}$ MeV \cite{Beringer:1900zz}. However,
the CLAS Collaboration at JLab  searched this state through the photoproduction  process $\gamma \, p\to \pi^+\pi^+\pi^- (n)$, without finding evidence of it  \cite{Mecking:2003zu}.

The third state with $J^{PC}=1^{-+}$, the meson $\pi_1(2015)$, has been reported by the E852 Collaboration in Refs. \cite{Kuhn:2004en,Lu:2004yn}.
A second peak in $\pi^- p$ interaction has been observed at about 2001 MeV in $f_1 \pi$ final state, which is  absent in the $\eta \pi$, $\eta^\prime \pi$ and $ \rho \pi$ final states, and this represents a new hybrid meson candidate \cite{Kuhn:2004en}. The signal has also been seen in $b_1(1235)\pi^-$ at about 2014 MeV \cite{Lu:2004yn}. 

On the basis  of the experimental evidences collected so far, and reviewed   in \cite{Ketzer:2012vn},
 the identification of these  three resonances is  a debated issue, and the  experiments at  BES III in Beijing, at the Jefferson Laboratory,   the PANDA experiment at GSI, LHCb and COMPASS at CERN and the new Super-B factory
at KeK will shed light on their actual existence and properties. 
If new hadronic states can be obtained explicitely considering the dynamics of gluonic excitations,
hybrid configurations with non-exotic quantum numbers should also be present in the meson spectrum. However, their identification  in this case is more difficult, due to the mixing with the ordinary $q \bar q$ configurations, and one has to mainly look for overpopulation of the levels with respect
to, e.g., the quark model predictions. Candidates for  hybrid  non-exotic resonances are  present in the charmonium system, namely  $Y(4260)$  with $J^{PC}=1^{--}$  \cite{Swanson:2006st}.

The properties of light and heavy hybrid mesons can  be studied through different theoretical approaches. Focusing on the lightest $1^{-+}$ meson, 
using  several phenomenological models some mass predictions  have been obtained, which are around 1.5 GeV in the old bag model \cite{Chanowitz:1982qj},  in the range 1.7-1.9 GeV in the flux-tube model \cite{Isgur:1984bm},  within 1.8-2.2 GeV in a model with  ``constituent''  gluon  \cite{Ishida:1991mx}.
Using  QCD sum rules, this mass is expected  in the range 1.5-1.8 GeV \cite{Balitsky:1986hf,Latorre:1985tg}, and  the value  $M=1.73^{+0.10}_{-0.09}$ GeV is found in Ref. \cite{steele}.
Lattice QCD analyses obtain  $M(\pi_{1^{-+}})=1.74 \pm 0.25$ GeV \cite{Hedditch:2005zf}  and $M\sim 2$ GeV \cite{Dudek:2011bn}. 
Hence, the predictions of the mass of the lightest $1^{-+}$ meson drastically   depend  on the theoretical approach.
With the exception of   the bag model,  a heavier mass with respect to the  $\pi_1(1400)$ and $\pi_1(1600)$  hybrid  candidates is generally predicted.

In the following,  we study the light $J^{PC}=1^{-+}$  hybrid states  in a framework  inspired by the gauge/string duality, using a holographic model of QCD  able to describe several features of the low-energy strong interaction phenomenology \cite{Karch:2006pv}. 
We shall also analyze the differences with the outcome of another holographic QCD model,  the so-called hard-wall model \cite{Kim:2008qh}. Moreover, we shall investigate the stability of the hybrid configurations against  thermal effects in comparison with other hadrons  with non-exotic quantum numbers.

\section{Model}
We adopt  the so-called
 soft-wall model,  introduced  as a phenomenological approach to low-energy QCD  inspired by the gauge/gravity correspondence \cite{Karch:2006pv}.
The model  aims at describing several nonperturbative features of QCD by  a  semiclassical theory living in a 5-dimensional anti-de Sitter space ($AdS_5$)   assumed to be dual to QCD.
The framework is inspired by the AdS/CFT correspondence, a conjecture   relating type IIB string theory living in $AdS_5 \times S^5$ ($S^5$ is a 5-dimensional sphere) with $\cal{N}$=4 Super-Yang Mills theory living in a 4-dimensional Minkowski space \cite{Maldacena:1997re}.
We use  Poincar\'e coordinates with  the line element  of the $AdS_5$ space written as
\be\label{eq:metric}
ds^2=\frac{R^2}{z^2}(dt^2-d\bar x^2-dz^2) \qquad\qquad z>0 \,;
\ee
 $R$ is the radius of the AdS space, $(t,\vec x)$ the ordinary four-dimensional coordinates,  and $z$ the fifth holographic coordinate.
Following the AdS/CFT correspondence dictionary  \cite{Witten:1998qj,Gubser:1998bc}, local gauge-invariant QCD operators with  conformal dimension $\Delta$ can be described in the dual theory by proper fields,
with  mass of a $p$-form field  fixed by the relation $m_5^2 R^2=(\Delta-p)(\Delta+p-4)$.
The field dynamics and interactions are described by an action containing a background dilaton field introduced to break conformal invariance.

To describe hybrid $1^{-+}$ mesons, we use the QCD local operator  $J_\mu^a=\bar q T^a G_{\mu\nu} \gamma^\nu q$, 
with $G_{\mu\nu}$ the gluon field strength and
$T^a$ flavour matrices normalized to  Tr$[T^aT^b]=\delta^{ab}/2$. 
This QCD operator  has a dual field which is a 1-form, $H_\mu=H_\mu^a T^a$, with mass  $m_5^2 R^2=8$.
The dynamics of such a field is described by the Proca-like action (with AdS background):
\be\label{eq:action}
S=\frac{1}{k}\int d^5x \, \sqrt{g}\, e^{-c^2z^2} \, \mbox{Tr}\left[-\frac{1}{4} F^{MN}F_{MN} +\frac{1}{2} m_5^2 H_M H^M  \right]\,, 
\ee
with $F^{MN}=\partial_M H_N-\partial_N H_M$, and $M, N$  5-dimensional indices.
$g$ is the determinant of the metric (\ref{eq:metric}) and $k$  a parameter introduced to make the action dimensionless.
The  $e^{-c^2z^2}$ factor  introduces  the mass parameter  $c$  breaking  the conformal symmetry.

The Euler-Lagrange equations for the field $H^N(x,z)$, stemming  from the action \eqref{eq:action}, are
\be\label{eq:EulerLag}
\partial_M\left( \sqrt{g}\, e^{-c^2z^2}\, F^{MN}\right)+\frac{8}{R^2} \sqrt{g}\, e^{-c^2z^2}\, H^N=0\,.
\ee
The $H_\mu$ field (greek letters indicate 4-dimensional indices) can be decomposed in a longitudinal $H^\mu_\parallel$, and  in a transverse $H^\mu_\perp$  component  which satisfies the condition $\partial_\mu H^\mu_\perp=0$ and can be used to describe the $1^{-+}$ mesons.
From Eq.~\eqref{eq:EulerLag}, the equation of motion for $H_\mu^\perp$ in the Fourier space is
\be
\partial_z\left( \frac{e^{-c^2z^2}}{z} \partial_z H_\mu^\perp\right)+q^2\, \frac{e^{-c^2z^2}}{z} H_\mu^\perp-8\, \frac{e^{-c^2z^2}}{z^3}\, H_\mu^\perp=0 \,,
\ee
with mass  spectrum and normalized eigenfunctions 
\bea\label{eq:massspectrum}
q^2_n&=&M_n^2=4 c^2(n+2)   \,\,\,\ ,  \\
H_n&=&\sqrt{\frac{2\, n!}{(n+3)!}} \, c^4\, z^{4}\, L_n^3(c^2z^2) \, 
\eea
($n$ integer  and $L_n^m$ the generalized  Laguerre polynomial). This spectrum can be compared with the one of mesons with different quantum numbers and quark content, collected in Table
\ref{tab:spectrum}.  In all cases, linear Regge trajectories $M_n^2 \simeq n$ with the same slope are obtained, which was the main motivation for  introducing the soft-wall \cite{Karch:2006pv}.

\begin{table}
\centering
\begin{tabular}{c  c c c c}
\hline
$J^{PC}$&$1^{--}$  $(q \bar q)$ \cite{Karch:2006pv} \,&$0^{++}$ $(q \bar q)$ \cite{Colangelo:2008us}&$0^{++}$ (glueball) \cite{Colangelo:2007pt} &$1^{-+}$ \\
\hline   
 $M^2_n$& $c^2(4 n+4)$&$c^2(4 n+6)$&$c^2(4 n+8)$&$c^2(4 n+8)$\\
\hline
  \end{tabular}
\caption{Mass spectrum of $q \bar q$, glueballs and hybrids predicted in the soft-wall model.}\label{tab:spectrum}
\end{table}

For  the lightest hybrid state the result is $M_0^2=8 c^2$ and setting $c$  from the  $\rho$ meson mass and the relation in Table \ref{tab:spectrum}, $c=M_\rho/2=388$ MeV, the value $M_0 \sim 1.1$ GeV is obtained.
Other choices of $c$ are possible.
For example, one can  fix $c$ including  the mass of the excited states of the $1^{--}$ spectrum, or the whole Regge trajectory, as done in \cite{Andreev:2006ct}. 
In the latter case, we  obtain $c\sim 474$ MeV and   $M_0\sim 1.34$ GeV. In any case, 
the value of the mass is  always lower than the preditions obtained by other approaches, and  favours $\pi_1(1400)$ rather than $\pi_1(1600)$ as the lightest $1^{-+}$ meson.
A similar conclusion  is drawn from the hard-wall model, in which the predicted mass of the lightest hybrid is $M_0=1476$ MeV \cite{Kim:2008qh}.
The mass of the radial excitations can also be determined: the mass of the first radial excitation, in the soft-wall model, is  given by $M_1/M_0= \sqrt{3/2}$, while $M_1/M_0=2.61/1.476$ is found  in the hard-wall model \cite{Kim:2008qh}.
The difference between the two determinations of the mass of the first excited state underlines the main difference between the two models,  the hard-wall model being characterized by the spectral condition  $M_n^2 \simeq n^2$.

\begin{figure}
 \centering
 \includegraphics[width=8cm]{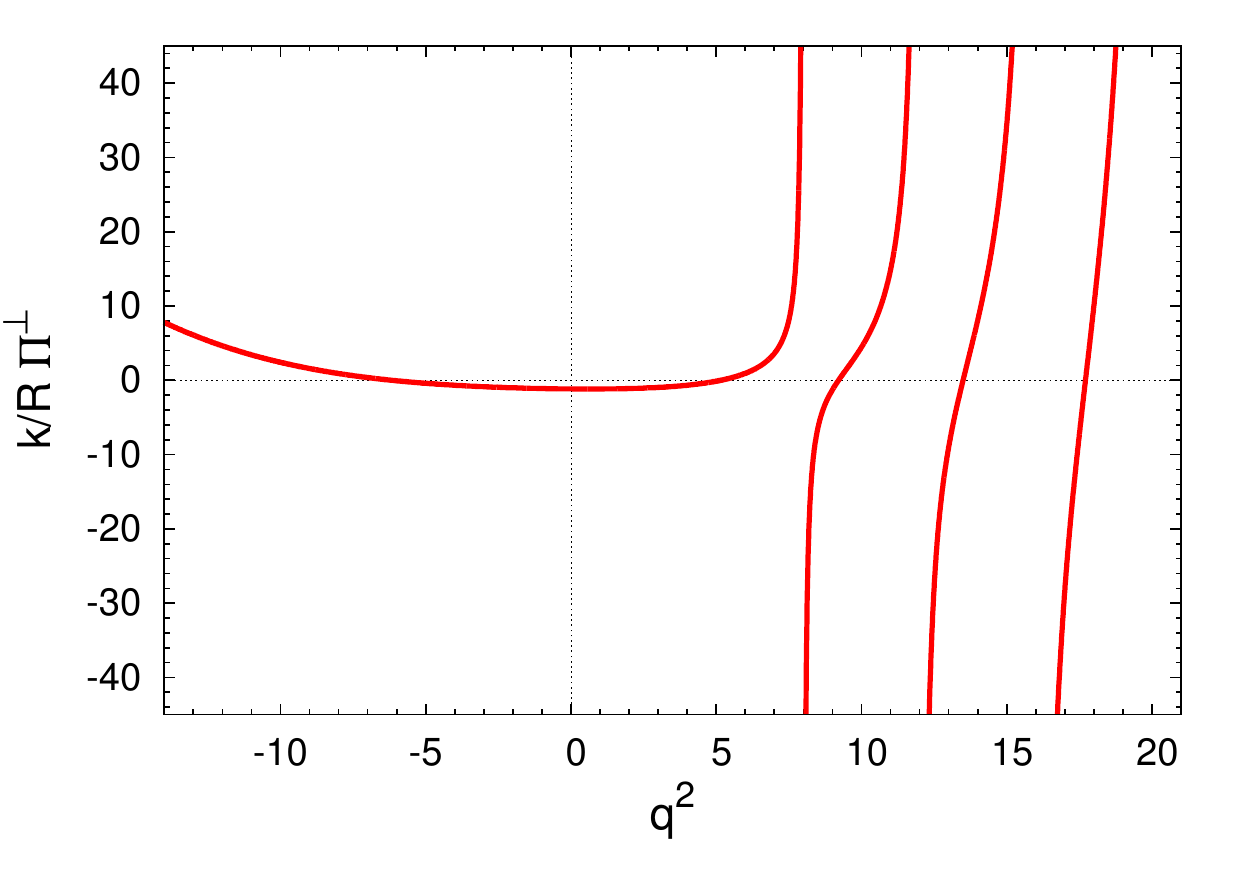}
 \caption{$\Pi^\perp(q^2)$ in Eq.\eqref{eq:Piperp} (with $\epsilon=1$).}
 \label{fig:Piperp}
\end{figure}

Other properties  can be extracted from the two-point correlation function
\be
\Pi^{ab}_{\mu\nu}= i \int d^4 x \, e^{i q x} \langle 0| T[J^a_\mu(x) J^b_\nu(0)]  | 0 \rangle  \,\,\, .
\ee
In the holographic approaches the computation of this quantity is based on the identification between the partition function of the conformal field theory and the one of the supergravity theory \cite{Witten:1998qj}.
This allows one to get the two-point correlation function of an operator by functional derivation of the supergravity on-shell action with respect to the source of the operator.
For the hybrid current operator, this relation can be written as
\be
\Pi_{\mu\nu}^{ab}(q^2)=\left. \frac{\delta^2 S_{os}}{\delta H_0^{a\mu} \delta H_0^{b\nu}}\right|_{H_0=0}\,,
\ee
where $H_0^\mu(q^2)$ is the source in the Fourier space,  related to the hybrid field   $H^\mu(z,q)=H(z,q^2)H_0^\mu(q^2)$ by  the bulk-to-boundary propagator  $H(z,q^2)$.
$\Pi_{\mu\nu}^{ab}(q^2)$ is  the  sum of a transverse and a longitudinal contribution:
\be
\Pi_{\mu\nu}^{ab}(q^2)=-\left(\eta_{\mu\nu}-\frac{q_\mu q_\nu}{q^2}\right) \frac{\delta^{ab}}{2}\Pi^\perp(q^2)+\frac{q_\mu q_\nu}{q^2} \frac{\delta^{ab}}{2} \Pi^\parallel(q^2)\,.
\ee 
$H^\perp(z,q^2)$ is determined as  the solution of  Eq. \eqref{eq:EulerLag} with  boundary condition at small $z$: $H^{\perp}(z,q^2)=1/z^2+\CMcal{O}(z^0)$,  and regular behaviour at $z\to\infty$:
\be
H^{\perp}(z,q^2)=\frac{1}{2}\Gamma \left(2-\frac{q^2}{4c^2} \right) \, z^{-2}\, U\left (-1-\frac{q^2}{4c^2},-2,c^2z^2\right)\,,
\ee
in terms of   the Tricomi confluent hypergeometric function $U(a,b,z)$. $\Pi^\perp(q^2)$ is given by:  $\Pi^\perp(q^2)=\left. \frac{R}{k} \frac{e^{-c^2z^2}}{z} H^\perp(z,q^2)\partial_z H^\perp(z,q^2)\right|_{z=0}$, so that
\begin{eqnarray}\label{eq:Piperp}
\Pi^{\perp}(q^2)&=&\frac{R}{k}\left(-\frac{2}{\epsilon^6} + \frac{c^2 - q^2/4}{\epsilon^4} + \frac{c^2 q^2}{4 \epsilon^2} + \frac{1}{1152}(-192 c^6 + 16 c^4 (-7 + 3 \gamma_E) q^2 + \right.\nonumber\\
&& \left. 36 c^2 q^4 + (7 - 3 \gamma_E) q^6 - 3 q^2 (-16 c^4 + q^4) (HN(1 - \frac{q^2}{4 c^2}) +  2 \log(c\epsilon)))\right)\,. 
\end{eqnarray}
$HN$ is the harmonic number function,  and  $\epsilon$  an inverse renormalization scale.
A plot of $\Pi^{\perp}(q^2)$ is shown in Fig.~\ref{fig:Piperp}, setting  $\epsilon=1$.
The poles of $\Pi^{\perp}(q^2)$ reproduce the mass spectrum \eqref{eq:massspectrum} with  residues 
\be
F_n^2=\frac{2Rc^8}{3k}(n+3)(n+2)(n+1)  \,\,\, .
\ee
The  parameter $k$ can be fixed  by matching the leading-order perturbative QCD expression for  $\Pi^\perp(Q^2)$  at $Q^2=-q^2\to\infty$ \cite{Balitsky:1986hf}
\be
\Pi_{\mbox{QCD}}^\perp(Q^2) = \frac{1}{960\pi^4} \log(Q^2) Q^6  \,\,\, 
\ee
with the asymptotic $Q^2 \to\infty$ expansion of Eq.~\eqref{eq:Piperp}:
\be
\Pi^\perp(Q^2) =\frac{R}{384k} \log(Q^2) Q^6 \,\,\, , 
\ee
 obtaining $R/k=2/(5\pi^4)$.
The results of this straightforward  analysis are, therefore:
\begin{itemize}
\item the soft-wall model produces linear Regge trajectories also for the hybrid $1^{-+}$ states,
\item a hierarchy with the vector and scalar glueball masses is found $M_\rho < M_G\simeq M_{1^{-+}}$,
\item the expression for the residues is  obtained,
\end{itemize}
which are interesting to compare with the outcome of  different approaches.

\section{Thermal effects on the hybrid meson spectrum}\label{sec:thermal}
The soft-wall model  allows a simple description of  finite temperature effects on  hadronic systems.
An interesting quantity to study is the spectral function. For the hybrid states, 
at zero temperature the spectral function is represented by an infinite number of delta functions  centered at the masses in  \eqref{eq:massspectrum}.
For vector mesons, scalar mesons and glueballs  it has been found that, increasing the temperature, these peaks broaden,  and their positions move towards lower values 
\footnote{We are not considering in this discussion the possibility of a Hawking-Page transition \cite{Colangelo:2009ra}. }\cite{Fujita:2009wc,Colangelo:2009ra}.
The analogous computation for hybrid mesons can be carried out.

\begin{figure}[t]
 \centering
 \includegraphics[width=8cm]{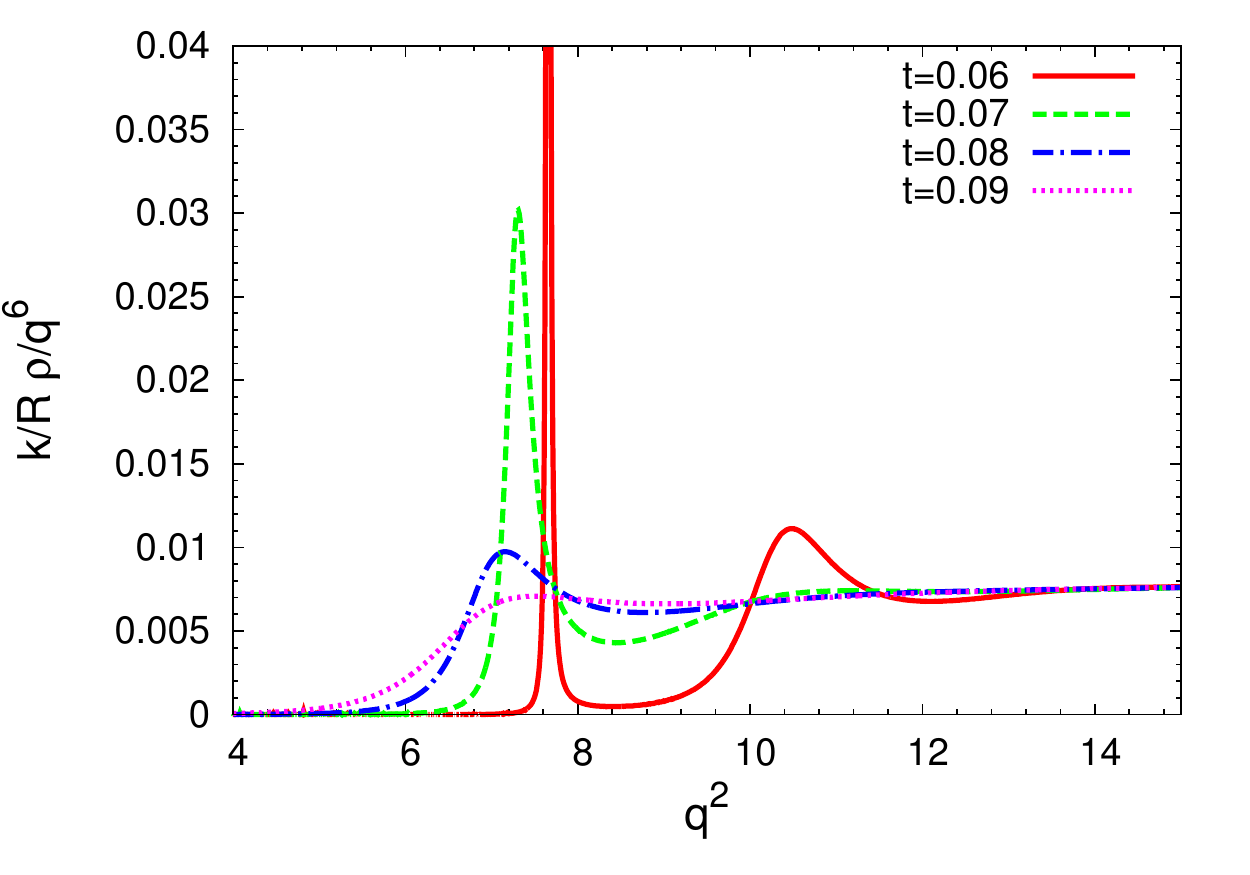}
  \includegraphics[width=8cm]{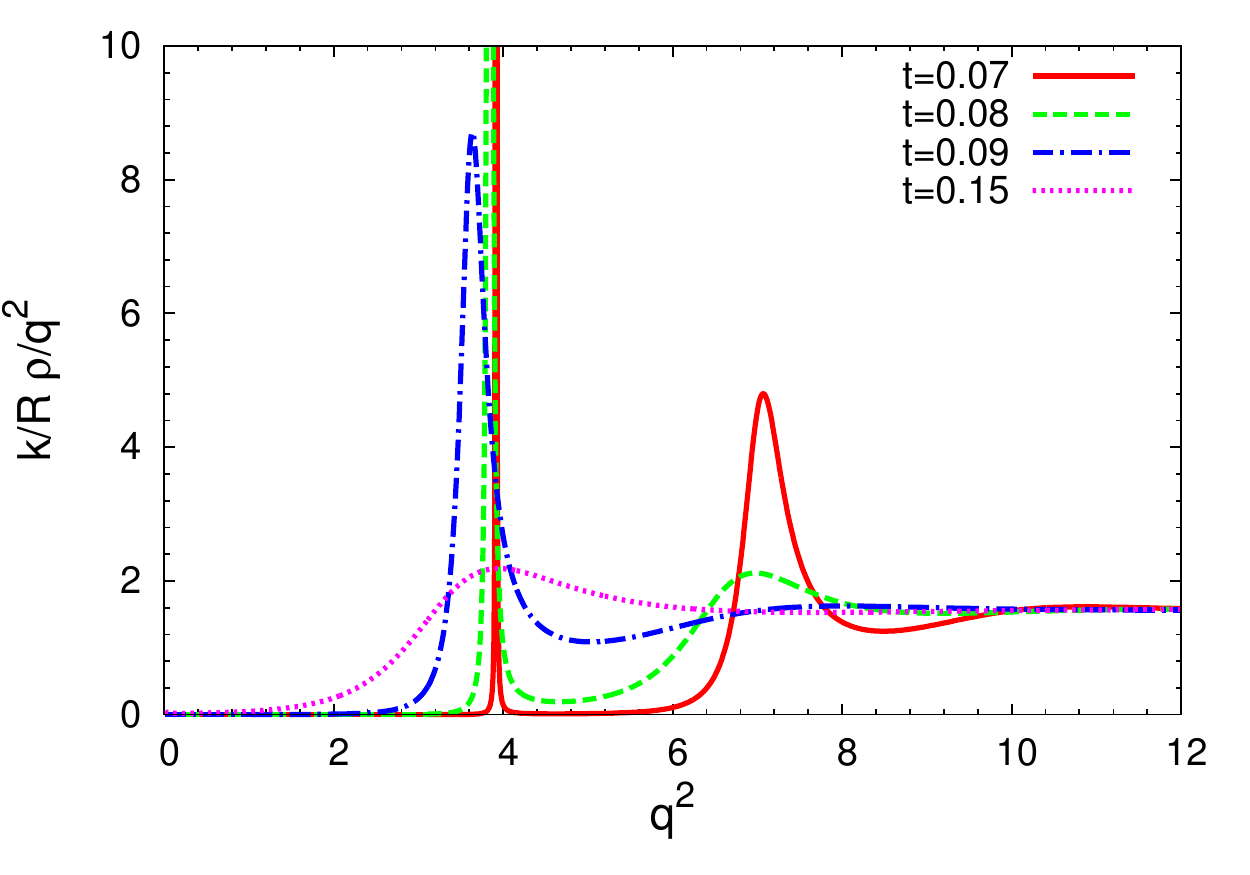}
 \caption{Left (right) panel: spectral function of  $1^{-+}$ ($1^{--}$) mesons at different values of temperature. The scale $c=1$ has been set.}
 \label{fig:spfunc}
\end{figure}

\begin{figure}
 \centering
 \includegraphics[width=7.cm]{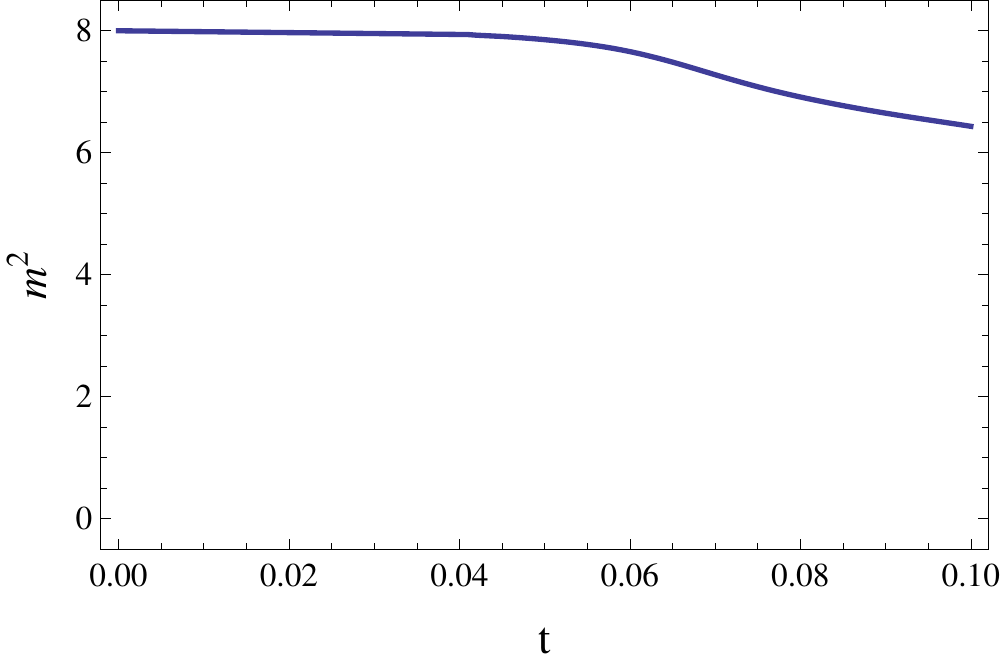} \hspace*{.4cm} 
 \includegraphics[width=7.cm]{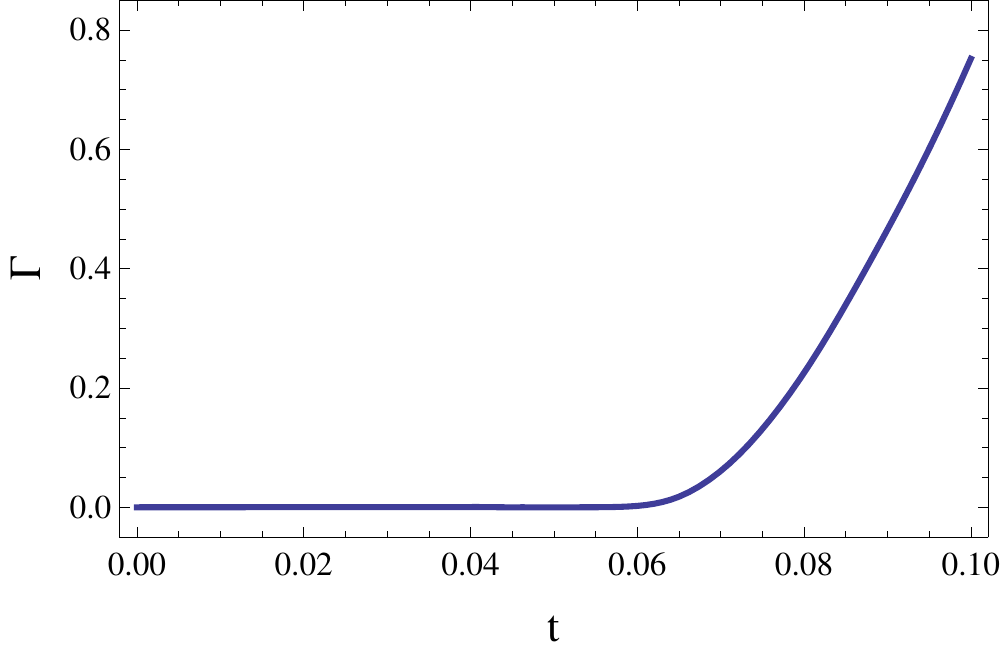}
 \caption{Squared mass (left) and width (right) of the lightest  $1^{-+}$ meson at increasing  temperature (with  $c=1$).}
 \label{fig:mass}
\end{figure}

Temperature can be incorporated in the holographic models  introducing a black hole in the $AdS$ space, with the position of its horizon $z_h$ related to the (inverse) temperature, $z_h=1/(\pi T)$.
Now, the line element is
\bea\label{eq:metricBH}
ds^2&=&\frac{R^2}{z^2}\left(f(z)dt^2-d\bar x^2-\frac{dz^2}{f(z)}\right)  \nn \\
 f(z)&=&1-z^4/z_h^4 \,, \,\,\quad\quad  0<z<z_h \,. 
\eea
From the action \eqref{eq:action},  using the metric in \eqref{eq:metricBH}, the equation of motion for a spatial component of the transverse field $H_i^\perp(z,q)$ can be obtained as in the $T$=0 case:
\be\label{eq:thermaleom}
\partial_z\left(\frac{e^{-c^2z^2} \, f(z)}{z}  \partial_z H_i^\perp(z,q) \right)+\frac{e^{-c^2z^2}}{z} \left(\frac{q_0^2}{f(z)}-\bar q^2 \right)H_i^\perp(z,q)-8 \frac{e^{-c^2z^2}}{z^3}H_i^\perp(z,q)=0 \,;
\ee
we restrict ourselves to the meson rest-frame  $\bar q$=0.
The bulk-to-boundary propagator is the solution of this equation with the same boundary condition at $z\to 0$ as in the $T=0$ case ($H^{\perp}(z,q^2)=1/z^2+\CMcal{O}(z^0)$), while near the black-hole horizon an  incoming-wave behavior is required,   $H^{\perp}(z,q^2)\sim (1-z/z_h)^{-i\sqrt{q^2}z_h/4} (1+\CMcal{O}(1-z/z_h))$,
to determine the retarded Green's function \cite{Son:2002sd}:
\be
\Pi_R(q^2)=\left. \frac{R}{k} \frac{e^{-c^2z^2} f(z)}{z} H^{\perp}(z,q^2)\partial_z H^{\perp}(z,q^2) \right|_{z\to0} \,.
\ee
The resulting spectral function,  the imaginary part of the retarded Green's function, is shown in Fig.~\ref{fig:spfunc}  for a few  values of temperature (in units of $c$: $t=T/c$).
The broadening of the peaks and the shift of their position towards lower values when the temperature is increased,  are clearly visible.  The temperature  dependence of the mass and width of the lowest lying state,
determined fitting the spectral function by a Breit-Wigner formula, 
\begin{equation}\label{breitwigner}
\rho(\omega^2)=\frac{a \, m \, \Gamma \, \omega^b}{(\omega^2-m^2)^2+m^2 \Gamma^2} 
\end{equation}
with parameters $a$ and $b$,
is depicted in Fig.~\ref{fig:mass}.
We determine the melting temperature 
$t\sim 0.10$ for the hybrid  meson looking  at the value of $t$ where the peak in the spectral function is reduced by a factor $\simeq 15-20$ with respect to the point where the width starts to broaden ($t \simeq 0.065$), a criterium already adopted in \cite{Colangelo:2009ra}. 

\begin{figure}[h!]
 \centering
 \includegraphics[width=8cm]{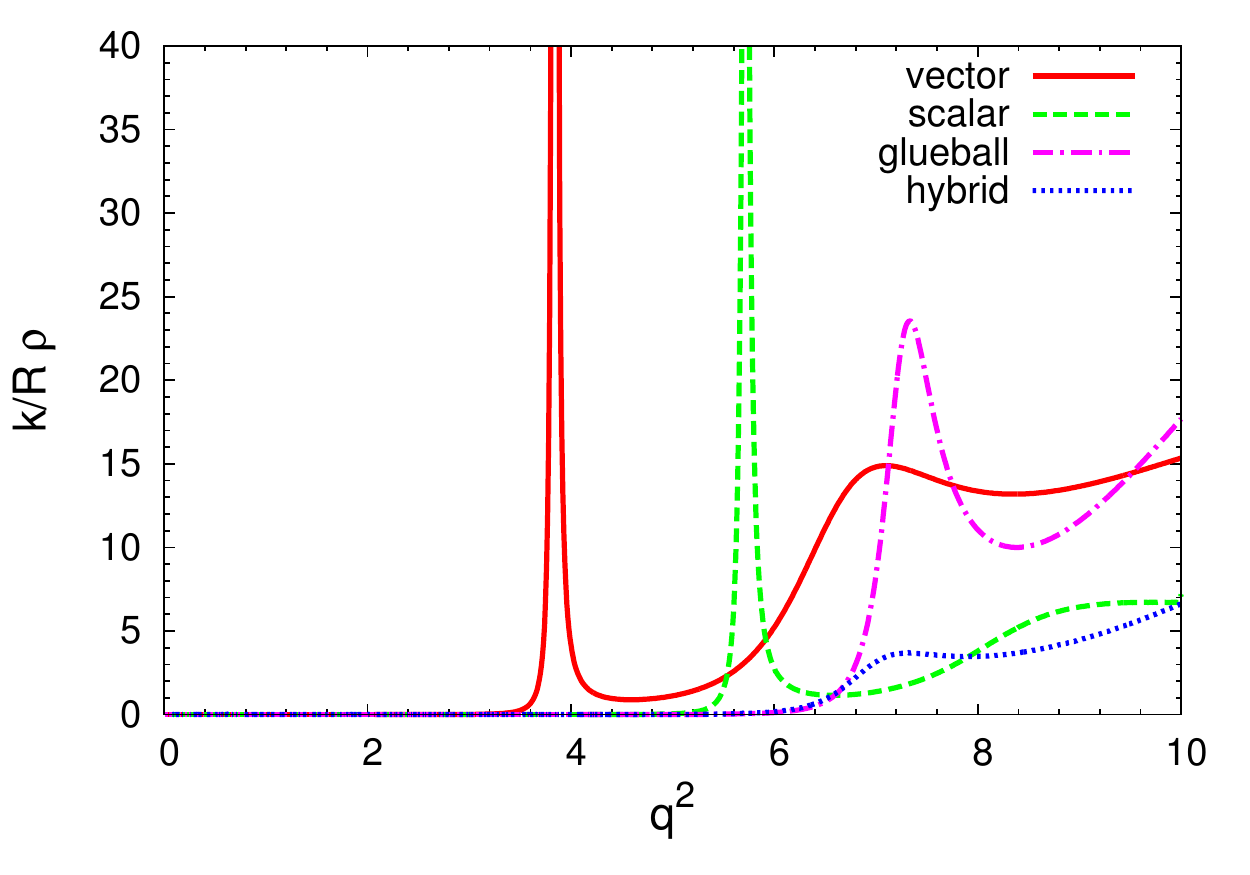}
 \caption{Spectral functions of vector mesons, scalar mesons, scalar glueballs and $1^{-+}$ hybrid mesons at $t=0.08$, setting the scale $c=1$.}
 \label{fig:sfT008}
\end{figure}

In  Fig.~\ref{fig:spfunc} we also plot, for comparison, the spectral function of vector mesons ($J^{PC}=1^{--}$). This shows  that, although the general behaviour of the spectral function with temperature is similar, hybrid states are more unstable than $1^{--}$ states and melt at lower temperatures. 
\begin{table}[b]
\centering
\begin{tabular}{c  c c c c}
\hline
$J^{PC}$&$1^{--}$  $(q \bar q)$ \cite{Mamani:2013ssa} \,&$0^{++}$ $(q \bar q)$ \cite{Colangelo:2009ra}&$0^{++}$ (glueball) \cite{Colangelo:2009ra}&$1^{-+}$ \\
\hline   
 $ t=T/c$& $0.23$&$0.18$&$0.12$&$0.10$\\
\hline
  \end{tabular}
\caption{Melting temperature of  $q \bar q$, glueball and hybrid states predicted in the soft-wall model.}\label{tab:temp}
\end{table}
 Vector mesons melt at $t\sim 0.23$ \cite{Mamani:2013ssa},   while scalar glueballs melt at $t\sim 0.12$  \cite{Colangelo:2009ra},   as specified in Table \ref{tab:temp}. 
The different response to temperature of the various hadronic states can be appreciated considering  Fig.~\ref{fig:sfT008}, in which the spectral functions of states with different quantum numbers are computed at the same temperature $t=0.08$.

\begin{figure}[h!]
 \centering
 \includegraphics[width=8cm]{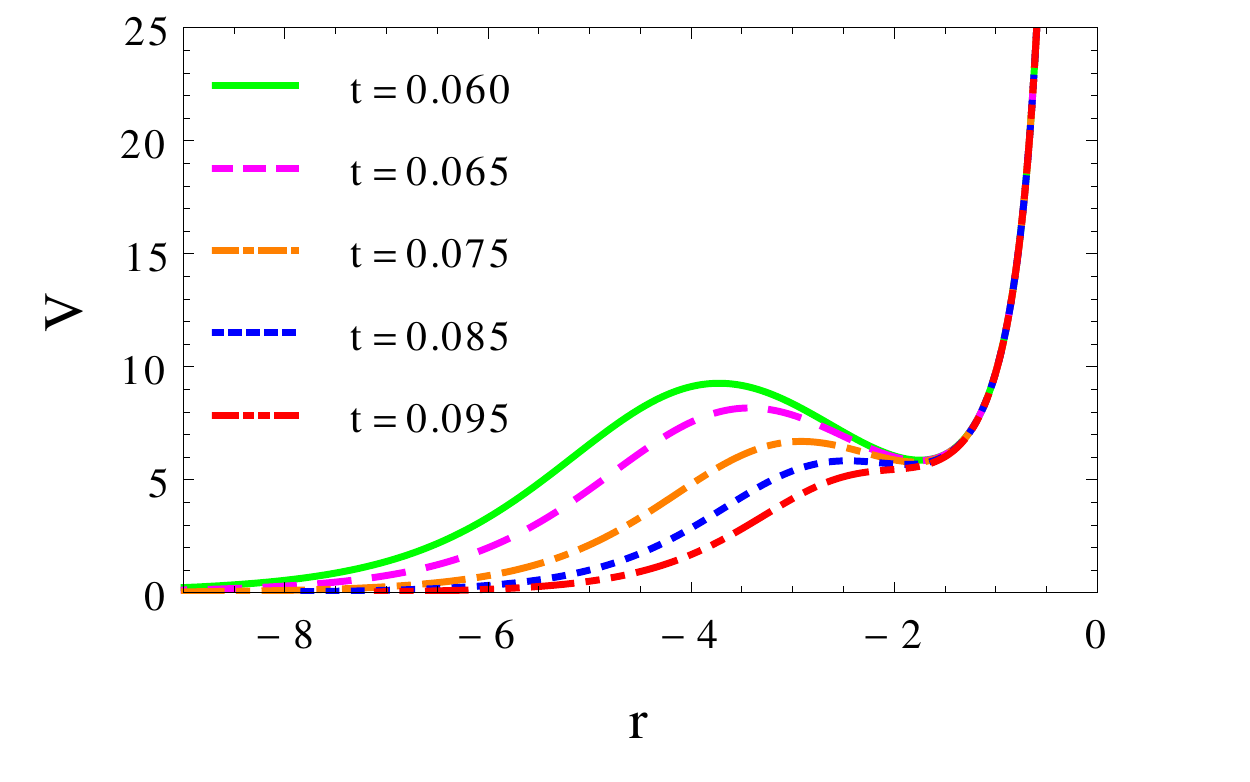}
 \caption{Potential $V$ vs the tortoise coordinate $r$ for several values of the temperature ($c$=1).}
 \label{fig:potential}
\end{figure}

The dissociation temperature of the hybrid mesons can be also determined  computing the potential  in  the Schr\"odinger-like equation to which  Eq.~\eqref{eq:thermaleom} can be transformed.
The tortoise coordinate $r$ can be defined, such that $\partial_r=-f(z)\partial_z$:
\be\label{eq:turtoise}
r=\frac{z_h}{2} \left(  -\arctan \left( \frac{z}{z_h}\right)+ \frac{1}{2} \ln\left(  \frac{z_h-z}{z_h+z}\right)    \right) \,\,\, . 
\ee
After a redefinition of the field $H^\perp=\sqrt{z} e^{c^2z^2/2}\psi$, Eq.~\eqref{eq:thermaleom}, with $\bar q=0$, can be written as
\begin{equation}
 -\psi''(r,q_0^2)+V(z(r)) \psi(r,q_0^2)=q_0^2 \psi(r,q_0^2)\,,
\end{equation}
with the potential 
\be
V(z)=\frac{f(z)}{z^2}\left(c^4 z^4 f(z)+\frac{4 c^2 z^6}{z_h^4}+\frac{35}{4}+\frac{5}{4} \frac{ z^4}{z_h^4}\right) 
\ee
in which the dependence on the temperature (through the horizon position) is explicit. 
The variation of the potential with temperature is shown in Fig.~\ref{fig:potential}. At a critical temperature the $r$ dependence of the potential becomes monotonous, so that no bound states can be formed.
Such a temperature coincides with the one inferred from the spectral function.

\section{Conclusions}
We have computed the spectrum and the decay constants of $1^{-+}$ hybrid mesons in the soft-wall holographic model of QCD.
For the lowest-lying  state we have found $M \sim 1.1-1.3$ GeV,  depending  on   the mass scale $c$. This value is close to the mass of $\pi_1(1400)$, the lightest one of the hybrid candidates.
The behaviour of $1^{-+}$ hybrid mesons at finite temperature has been studied,   showing that  dissociation occurs at a much lower temperature than for $1^{--}$ mesons, and also at a lower temperature  than in the case of other kind of hadrons.
This result can be interpreted as an indication that hybrid quark-gluon configurations, although present in the meson spectrum, suffer of larger instabilities with respect to the conventional configurations, and this could explain the difficulty in their detection in actual conditions, i.e. both in the finite number of colour condition ($N_c=3$), which is different from the holographic one holding in the large $N_c$ limit, and in the experimental environment.

\section*{Acknowledgments}
We thank M. Battaglieri, F. De Fazio and S. Nicotri  for fruitful discussions.


\newpage
\begin{thebibliography}{99}

\bibitem{early}
For a  review see
  C.~Amsler and N.~A.~Tornqvist,
  Phys.\ Rept.\  {\bf 389}, 61 (2004).

\bibitem{Swanson:2006st} 
  E.~S.~Swanson,
  Phys.\ Rept.\  {\bf 429}, 243 (2006).

\bibitem{Thompson:1997bs} 
  D.~R.~Thompson {\it et al.}  [E852 Collaboration],
  Phys.\ Rev.\ Lett.\  {\bf 79}, 1630 (1997).
  
\bibitem{Chung:1999we} 
  S.~U.~Chung {\it et al.}  [E852 Collaboration],
  Phys.\ Rev.\ D {\bf 60}, 092001 (1999).
  
\bibitem{Abele:1998gn} 
  A.~Abele {\it et al.}  [Crystal Barrel Collaboration],
  Phys.\ Lett.\ B {\bf 423}, 175 (1998).
  
\bibitem{Abele:1999tf} 
  A.~Abele {\it et al.}  [Crystal Barrel Collaboration],
  Phys.\ Lett.\ B {\bf 446}, 349 (1999).
  
\bibitem{Salvini:2004gz} 
  P.~Salvini {\it et al.}  [OBELIX Collaboration],
  Eur.\ Phys.\ J.\ C {\bf 35}, 21 (2004).

\bibitem{Adams:2006sa} 
  G.~S.~Adams {\it et al.}  [E862 Collaboration],
  Phys.\ Lett.\ B {\bf 657}, 27 (2007).
    
\bibitem{Beringer:1900zz} 
  J.~Beringer {\it et al.}  [Particle Data Group Collaboration],
  Phys.\ Rev.\ D {\bf 86}, 010001 (2012).

\bibitem{Adams:1998ff} 
  G.~S.~Adams {\it et al.}  [E852 Collaboration],
  Phys.\ Rev.\ Lett.\  {\bf 81}, 5760 (1998).
  
\bibitem{Ivanov:2001rv} 
  E.~I.~Ivanov {\it et al.}  [E852 Collaboration],
  Phys.\ Rev.\ Lett.\  {\bf 86}, 3977 (2001).
  
\bibitem{Kuhn:2004en} 
  J.~Kuhn {\it et al.}  [E852 Collaboration],
  Phys.\ Lett.\ B {\bf 595}, 109 (2004).
  
\bibitem{Lu:2004yn} 
  M.~Lu {\it et al.}  [E852 Collaboration],
  Phys.\ Rev.\ Lett.\  {\bf 94}, 032002 (2005).
  
\bibitem{Alekseev:2009aa} 
  M.~Alekseev {\it et al.}  [COMPASS Collaboration],
  Phys.\ Rev.\ Lett.\  {\bf 104}, 241803 (2010).

\bibitem{Mecking:2003zu} 
  B.~A.~Mecking {\it et al.}  [CLAS Collaboration],
  Nucl.\ Instrum.\ Meth.\ A {\bf 503}, 513 (2003).
  
\bibitem{Ketzer:2012vn} 
  B.~Ketzer,
  PoS QNP {\bf 2012}, 025 (2012);
  D.~Bettoni,
  PoS BORMIO {\bf 2012}, 004 (2012).
  
\bibitem{Chanowitz:1982qj} 
  M.~S.~Chanowitz and S.~R.~Sharpe,
  Nucl.\ Phys.\ B {\bf 222}, 211 (1983)
  [Erratum-ibid.\ B {\bf 228}, 588 (1983)];
  T.~Barnes, F.~E.~Close, F.~de Viron and J.~Weyers,
  Nucl.\ Phys.\ B {\bf 224}, 241 (1983).
  
\bibitem{Isgur:1984bm} 
  N.~Isgur and J.~E.~Paton,
  Phys.\ Rev.\ D {\bf 31}, 2910 (1985);
  T.~Barnes, F.~E.~Close and E.~S.~Swanson,
  Phys.\ Rev.\ D {\bf 52}, 5242 (1995).
  
\bibitem{Ishida:1991mx} 
  S.~Ishida, H.~Sawazaki, M.~Oda and K.~Yamada,
  Phys.\ Rev.\ D {\bf 47}, 179 (1993);
  I.~J.~General, S.~R.~Cotanch and F.~J.~Llanes-Estrada,
  Eur.\ Phys.\ J.\ C {\bf 51}, 347 (2007).
  
\bibitem{Balitsky:1986hf} 
  I.~I.~Balitsky, D.~Diakonov and A.~V.~Yung,
  Z.\ Phys.\ C {\bf 33}, 265 (1986);
  Phys.\ Lett.\ B {\bf 112}, 71 (1982).
  
\bibitem{Latorre:1985tg} 
  J.~I.~Latorre, P.~Pascual and S.~Narison,
  Z.\ Phys.\ C {\bf 34}, 347 (1987);
  K.~G.~Chetyrkin and S.~Narison,
  Phys.\ Lett.\ B {\bf 485}, 145 (2000);
  S.~Narison,
  Phys.\ Lett.\ B {\bf 675}, 319 (2009).

 \bibitem{steele}  
  Z.~-f.~Zhang, H.~-y.~Jin and T.~G.~Steele,
  arXiv:1312.5432 [hep-ph].
  
\bibitem{Hedditch:2005zf} 
  J.~N.~Hedditch, W.~Kamleh, B.~G.~Lasscock, D.~B.~Leinweber, A.~G.~Williams and J.~M.~Zanotti,
  Phys.\ Rev.\ D {\bf 72}, 114507 (2005).
  
\bibitem{Dudek:2011bn} 
  J.~J.~Dudek,
  Phys.\ Rev.\ D {\bf 84}, 074023 (2011).
   
\bibitem{Karch:2006pv} 
  A.~Karch, E.~Katz, D.~T.~Son and M.~A.~Stephanov,
  Phys.\ Rev.\ D {\bf 74}, 015005 (2006).
 
\bibitem{Kim:2008qh} 
  H.~-C.~Kim and Y.~Kim,
  JHEP {\bf 0901}, 034 (2009).

\bibitem{Maldacena:1997re} 
  J.~M.~Maldacena,
  Adv.\ Theor.\ Math.\ Phys.\  {\bf 2}, 231 (1998).

\bibitem{Witten:1998qj} 
  E.~Witten,
  Adv.\ Theor.\ Math.\ Phys.\  {\bf 2}, 253 (1998).
  
\bibitem{Gubser:1998bc} 
  S.~S.~Gubser, I.~R.~Klebanov and A.~M.~Polyakov,
  Phys.\ Lett.\ B {\bf 428}, 105 (1998).

\bibitem{Colangelo:2008us} 
  P.~Colangelo, F.~De Fazio, F.~Giannuzzi, F.~Jugeau and S.~Nicotri,
  Phys.\ Rev.\ D {\bf 78}, 055009 (2008).

\bibitem{Colangelo:2007pt} 
  P.~Colangelo, F.~De Fazio, F.~Jugeau and S.~Nicotri,
  Phys.\ Lett.\ B {\bf 652}, 73 (2007).
  
\bibitem{Andreev:2006ct} 
  O.~Andreev and V.~I.~Zakharov,
  Phys.\ Rev.\ D {\bf 74}, 025023 (2006).
  
\bibitem{Fujita:2009wc} 
  M.~Fujita, K.~Fukushima, T.~Misumi and M.~Murata,
  Phys.\ Rev.\ D {\bf 80}, 035001 (2009).
  
\bibitem{Colangelo:2009ra} 
  P.~Colangelo, F.~Giannuzzi and S.~Nicotri,
  Phys.\ Rev.\ D {\bf 80}, 094019 (2009).
 
\bibitem{Son:2002sd} 
  D.~T.~Son and A.~O.~Starinets,
  JHEP {\bf 0209}, 042 (2002).
    
\bibitem{Mamani:2013ssa}
  L.~A.~H.~Mamani, A.~S.~Miranda, H.~Boschi-Filho and N.~R.~F.~Braga,
  arXiv:1312.3815 [hep-th];
  A.~S.~Miranda, C.~A.~Ballon Bayona, H.~Boschi-Filho and N.~R.~F.~Braga,
  JHEP {\bf 0911}, 119 (2009).




\end{thebibliography}
\end{document}